# Plasticity of Zr-Nb-Ti-Ta-Hf high-entropy alloys

M. Feuerbacher, M. Heidelmann, C. Thomas

*Institut für Mikrostrukturforschung PGI-5, Forschungszentrum Jülich GmbH, 52425 Jülich, Germany*

**Abstract**

We have investigated the plastic deformation properties of non-equiatomic single phase Zr-Nb-Ti-Ta-Hf high-entropy alloys from room temperature up to 300 °C. Uniaxial deformation tests at a constant strain rate of $10^{-4}$ s$^{-1}$ were performed including incremental tests such as stress-relaxations, strain-rate- and temperature changes in order to determine the thermodynamic activation parameters of the deformation process. The microstructure of deformed samples was characterized by transmission electron microscopy. The strength of the investigated Zr-Nb-Ti-Ta-Hf phase is not as high as the values frequently reported for high-entropy alloys in other systems. We find an activation enthalpy of about 1 eV and a stress dependent activation volume between 0.5 and 2 nm$^3$. The measurement of the activation parameters at higher temperatures is affected by structural changes evolving in the material during plastic deformation.

## 1. Introduction

High-entropy alloys (HEAs) are metallic solid solutions composed of multiple principal elements, i.e. without a dominating basic component [1, 2]. Usually, alloys with 4 to 9, occasionally up to 20 components are considered. HEAs solidify in a simple average crystal structure, usually body-centered or face-centered cubic.

In a multicomponent alloy, the total free energy besides through the weighted contributions of the constituting elements is determined by the free energy of mixing, $\Delta G_{mix} = \Delta H_{mix} - T\Delta S_{mix}$ where $\Delta H_{mix}$ is the mixing enthalpy and $\Delta S_{mix}$ the mixing entropy. In an equimolar or near equimolar composition the mixing entropy can become dominant, stabilizing a random solid solution, if the elements are chosen such that the mixing enthalpy is neither too high positive (leading to segregation) nor too high negative (leading to the formation of ordered structures). In the ideal case the resulting structure is a perfectly disordered solid solution on a crystal lattice. HEAs are thus related to metallic glasses, but in contrast to the latter, they are thermodynamically stable, in particular at high temperatures.

HEAs were soon recognized as technologically relevant materials due to their high wear and oxidation resistance [3], high corrosion resistance [4] and fatigue endurance limit [5], their potential use as diffusion barriers [6] and soft magnetic materials [7], and their attractive mechanical properties. HEAs show a high hardness [1,8] and an appealing combination of high strength and good ductility [16]. The high hardness and strength is commonly ascribed to solid-solution strengthening, which is argued to be operative in an extreme form due to the high concentration of solute atoms [1, 9, 11].

In the present paper we address the intrinsic plastic properties of a HEA. We have characterized the plastic behavior of single phased, near single crystalline samples in the system Zr-Nb-Ti-Ta-Hf. We have measured stress-strain curves, determined thermodynamic activation parameters of the plastic deformation mechanism and provide an analysis of the deformed material by means of transmission electron microscopy (TEM) and scanning transmission electron microscopy (STEM).

## 2. Experimental

A composition of high-purity raw materials 12.7 at.% Zr, 30.8 at.% Nb, 17.7 at.% Ti, 30.8 at.% Ta, 8.0 at.% Hf [12] was arc molten and cast into a cylindrical copper mold. The so produced rod was recast in a zone-melting setup at a feed rate of 5 mm/h at about 2300 °C under 400 mbar Argon atmosphere. The final ingot consisted of a single body-centered phase with single grain sizes between about 0.5 and 1 mm.

Rectangular cuboidal samples of 1.4 x 1.4 x 3.2 mm$^3$ were cut from the ingot using a wire saw, their surfaces were ground on 1200 grit SiC paper. Plastic deformation experiments were performed in compression along the long sample axis in a modified ZWICK Z050 uniaxial testing system at constant true strain rate under closed-loop control. The strain was measured directly at the sample by a linear inductive differential transducer at an accuracy of ±10 nm.

Deformation experiments were performed between room temperature (RT) and 300 °C at a strain rate of $10^{-4}$ s$^{-1}$. The deformation temperatures around RT were controlled by air conditioning the ambient temperature, and at elevated temperatures by a cylindrical furnace equipped with an active temperature-gradient control. During the deformation experiments we have performed temperature-cycling tests, stress-relaxation tests and strain-rate changes in order to determine the thermodynamic activation parameters of the deformation process.

The activation volume V was determined by means of stress-relaxation tests, where the total strain is kept constant and the decrease of the stress is measured as a function of time. The data is evaluated according to [13]

$$V = \frac{kT}{m_s} \frac{\partial \ln(\dot{\sigma})}{\partial \sigma}\bigg|_T , \qquad (1)$$

where $m_s$ is the Schmid factor, k is Boltzmann's constant, $T$ is the deformation temperature, and $\sigma$ and $\dot{\sigma}$ are the applied stress and stress rate, respectively. The activation enthalpy $\Delta H$ was determined from temperature cycling experiments according to [13]

$$\Delta H = -kT^2 \left.\frac{\partial \ln(\dot{\varepsilon})}{\partial \sigma}\right|_T \left.\frac{\partial \sigma}{\partial T}\right|_{\dot{\varepsilon}} = -m_s TV \left.\frac{\partial \sigma}{\partial T}\right|_{\dot{\varepsilon}}. \tag{2}$$

Specimens for microstructural characterization were prepared by means of a dual beam FEI Helios Nanolab 400S focused ion beam apparatus from the centres of the deformation samples and from undeformed reference samples. Bragg-contrast TEM was carried out using a Philips CM20 transmission electron microscope operated at 200 kV, and STEM work using a probe corrected FEI Titan 80 – 300 operated at 300 kV and equipped with a high-angle annular dark-field (HAADF) detector.

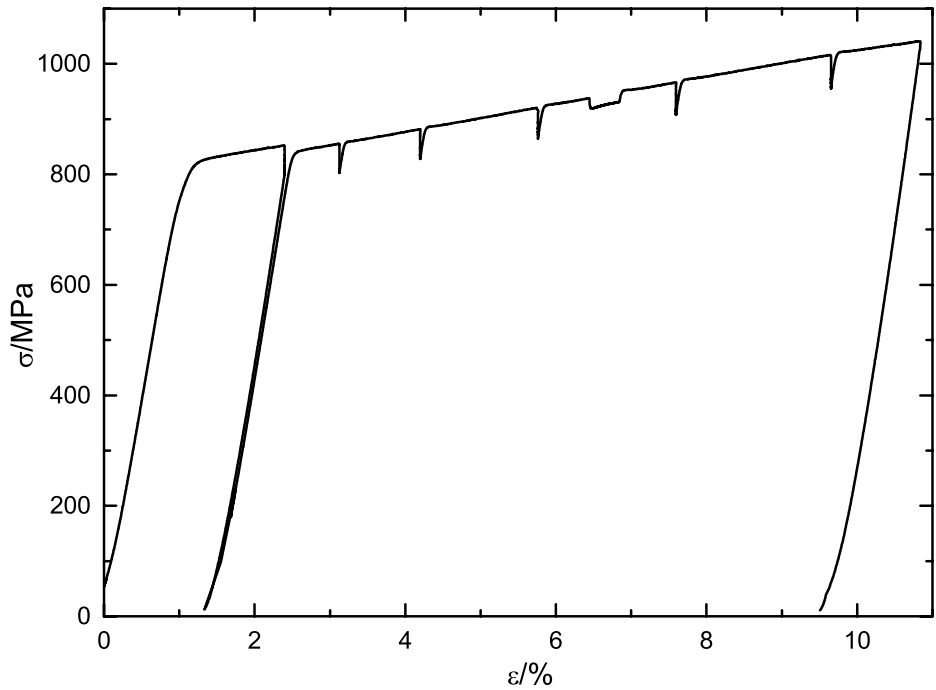

Fig. 1: True stress-true strain curve of the Zr-Nb-Ti-Ta-Hf high-entropy alloy at room temperature including six stress-relaxation tests (sharp dips), a temperature change at 2 % strain, and strain-rate changes at about 7 % strain.

## 3. Results

Fig. 1 shows a typical true stress-true strain curve at RT. The experiment is started at a deformation temperature of 19.6 °C. The sample deforms elastically with an apparent Young's modulus of 79 GPa up to a strain of about 0.8 %. The 0.2% yield stress $\sigma_{0.2}$, i.e. the stress at which the strain deviates by more than 0.2 % from the linear elastic behavior, amounts to 822 MPa. The curve continuously develops into a linear flow stress with a hardening rate of about 2.4 GPa without any yield point effect. At 2.4 % we halted the crosshead for a stress-relaxation test of two minutes. Subsequently the sample was unloaded, the temperature was increased to 24.7 °C and after thermal equilibration the sample was reloaded. As for initial loading the curve continuously develops toward a linear flow stress. The hardening rate remains unchanged at 2.4 GPa. The increase of deformation temperature results in a reduction of the flow stress by 12 MPa. At 3.1 % another stress-relaxation test is carried out. Deformation is then continued at the same temperature to strain of 10.8 %, at which the sample is unloaded. Four more stress-relaxation tests and one strain-rate change (from $10^{-4}$ s$^{-1}$ to $2 \cdot 10^{-4}$ s$^{-1}$ and back) were carried out.

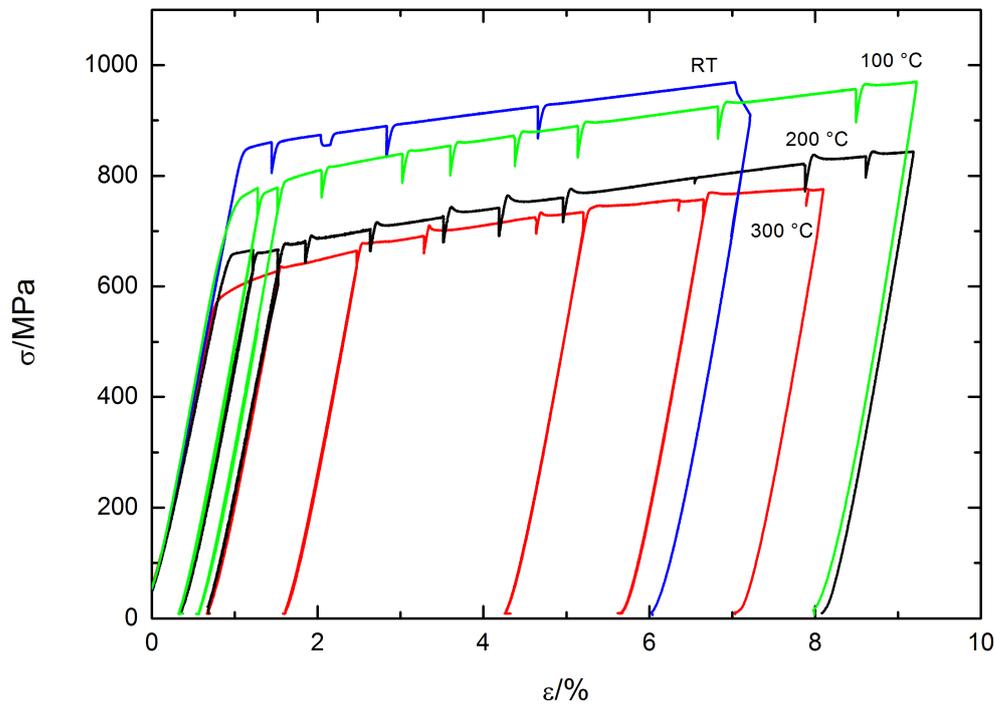

Fig. 2: True stress-true strain curves at temperatures between room temperature and 300 °C.

Fig. 2 shows true stress-true strain curves at temperatures between RT and 300 °C. The curve at RT is from a different run than that shown in Fig. 1 and reflects the reproducibility of the experiments. The flow stress (measured at a strain of 2 %) is higher by 28 MPa and the hardening lower by about 0.5 GPa, corresponding to relative errors of about 4 % and 20 %, respectively.

The flow stress decreases with increasing deformation temperature (see also Fig. 3) while the apparent Young's modulus and the hardening rate are essentially deformation-temperature independent. At all temperatures stress-relaxation tests and temperature cycling tests were carried out. For the curves at 60 °C (not shown) to 300 °C, the temperature was first increased by 10 °C and after a stress-relaxation at the increased temperature cycled back to the initial temperature. Please note that at 300 °C the overall shape of the curve differs from that at lower temperatures, and the temperature increase leads to a slight increase in flow stress. At this temperature a second temperature cycle around 5 % was carried out.

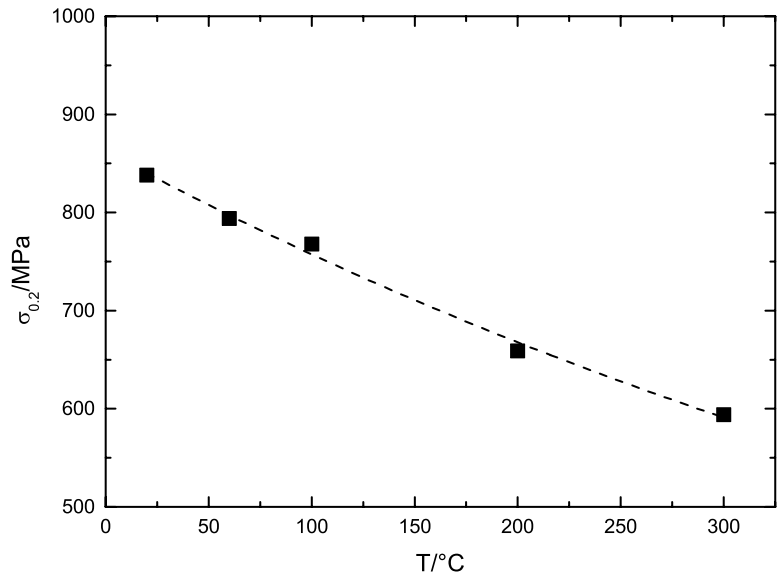

Fig. 3: Temperature dependence of the 0.2% yield stress $\sigma_{0.2}$.

Fig. 3 shows the temperature dependence of the 0.2% yield stress $\sigma_{0.2}$. For the temperatures at which more than one deformation experiment was carried out, the average value is shown. The line is a guide to the eye. The values of $\sigma_{0.2}$ drop from about 840 MPa at RT to 590 MPa at 300 °C.

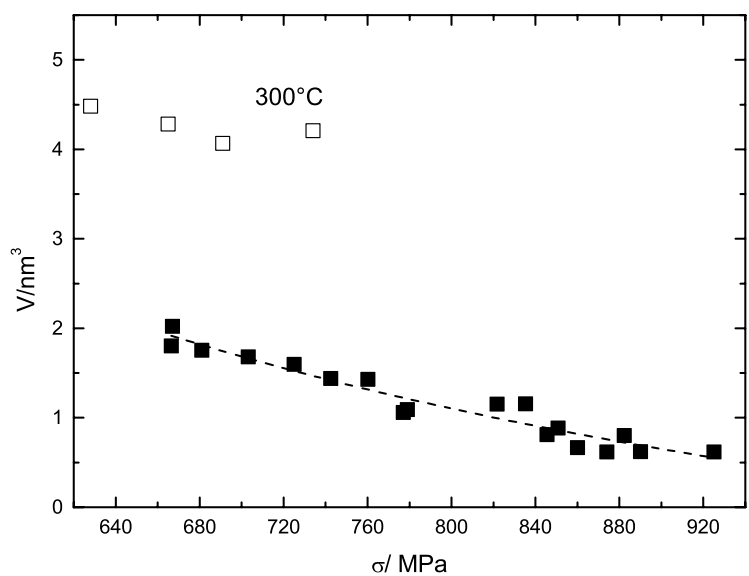

Fig. 4: Stress dependence of the activation volume V for temperatures below 200 °C (solid squares) and 300 °C (open squares). The line is a guide to the eye.

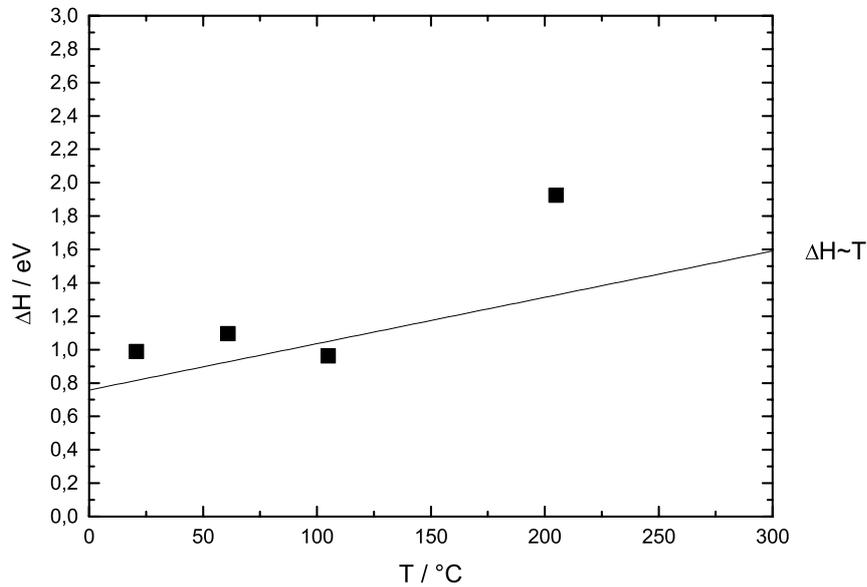

Fig 5: Temperature dependence of the activation enthalpy ΔH. The line represents a linear temperature dependence with ΔH(T=0) = 0 eV

Fig. 4 depicts the stress dependence of the activation volume determined according to Eq.(1) using a Schmid factor $m_s$ of 0.5. The values for deformation temperatures below 200 °C (solid squares) follow a single curve and decrease from about 2 nm$^3$ at 650 MPa to about 0.5 nm$^3$ at 920 MPa (dashed line). The data was obtained at different strains, ranging from 1.2 to 8.6 %. The activation volume in this temperature range is essentially independent of strain and deformation temperature and mainly depends on stress. The values determined at a deformation temperature of 300 °C (open squares) are considerably larger, ranging between about 4 and 4.5 nm$^3$, with a slight negative stress dependence.

Fig. 5 shows the temperature dependence of the activation enthalpy ΔH determined according to Eq.(2). At temperatures below 100 °C the values amount to about 1 eV with a scatter of about ±0.1 eV and lie close to the curve representing a linear temperature dependence (line) with ΔH(T=0) = 0 eV [14]. At 200 °C the determined value is significantly higher and at 300 °C no activation enthalpy can be determined from the experimental data since we have measured an increase of the flow stress resulting from an increase in deformation temperature. This would lead to a negative activation enthalpy, which has to be considered unphysical.

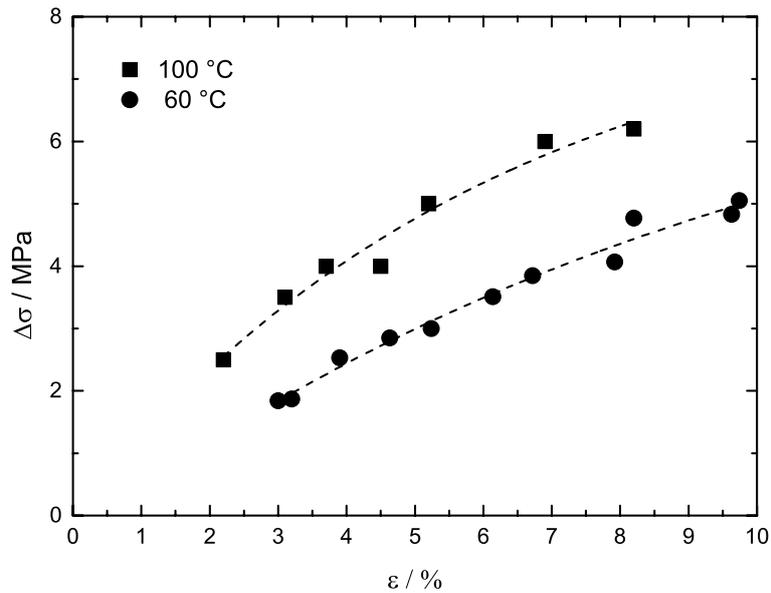

Fig. 6: Strain dependence of the stress overshoots for the deformation temperatures 60 and 100 °C (the lines are guides to the eye).

Careful inspection of Fig. 1 reveals that at higher temperatures stress-relaxations lead to a subsequent yield-point effect, i.e. an overshoot of the flow stress upon reloading. This effect occurs at deformation temperatures of 60 °C and higher. We have evaluated the amount of stress overshoot $\Delta\sigma$ by interpolating the flow stress over the stress relaxation and determining the difference between the maximum stress in the overshoot and the interpolated stress. Fig. 6 shows the strain dependence of the stress overshoots for the deformation temperatures 60 and 100 °C (the lines are guides to the eye). The overshoots are larger for higher deformation temperature and increase with increasing strain. At both temperatures an almost linear increase with a slight tendency towards saturation at high strains is found. The overshoots were also evaluated for the higher deformation temperatures of 200 and 300 °C. At these temperatures generally higher values, ranging between about 8 and up to 20 MPa, are found. However there is no clear strain dependence. The values irregularly scatter without an apparent tendency.

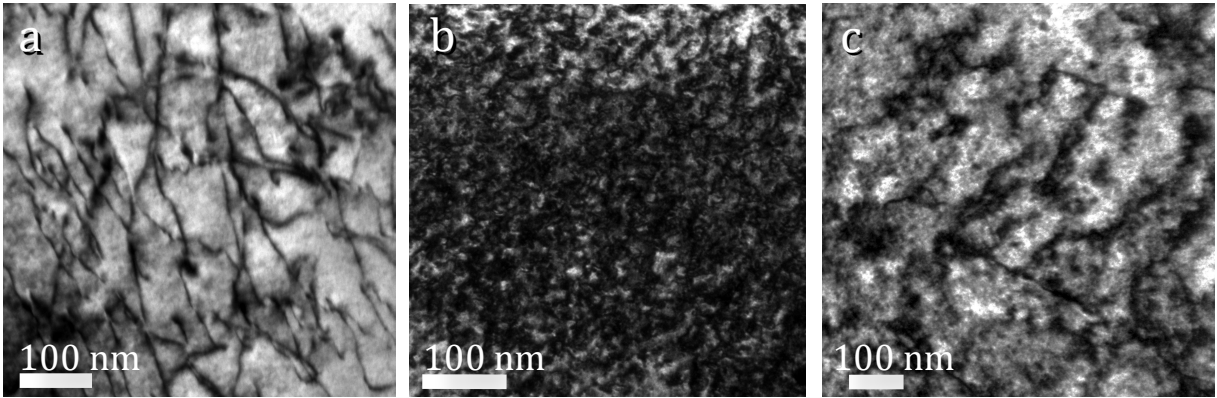

Fig. 7: Bright-field Bragg-contrast TEM micrographs. (a) as-grown material, (b) deformed at room temperature up to about 7 %, (c) deformed at 200 °C up to 9 %.

Fig 7 depicts a series of bright-field Bragg-contrast TEM micrographs. Fig. 7a was taken using a $(0\ 0\ \bar{2})$ reflection and depicts the typical microstructure of the as-grown material. The dislocation density, estimated by counting dislocation penetration points with the specimen surface, amounts to about $10^{10}$ cm$^{-2}$, and thus is already high in the undeformed material. Individual dislocations were analyzed by contrast-extinction experiments [15], which revealed that the dislocations have Burgers vectors parallel to the [1 1 1] direction. The micrograph in Fig 7 b was taken from a sample deformed at RT

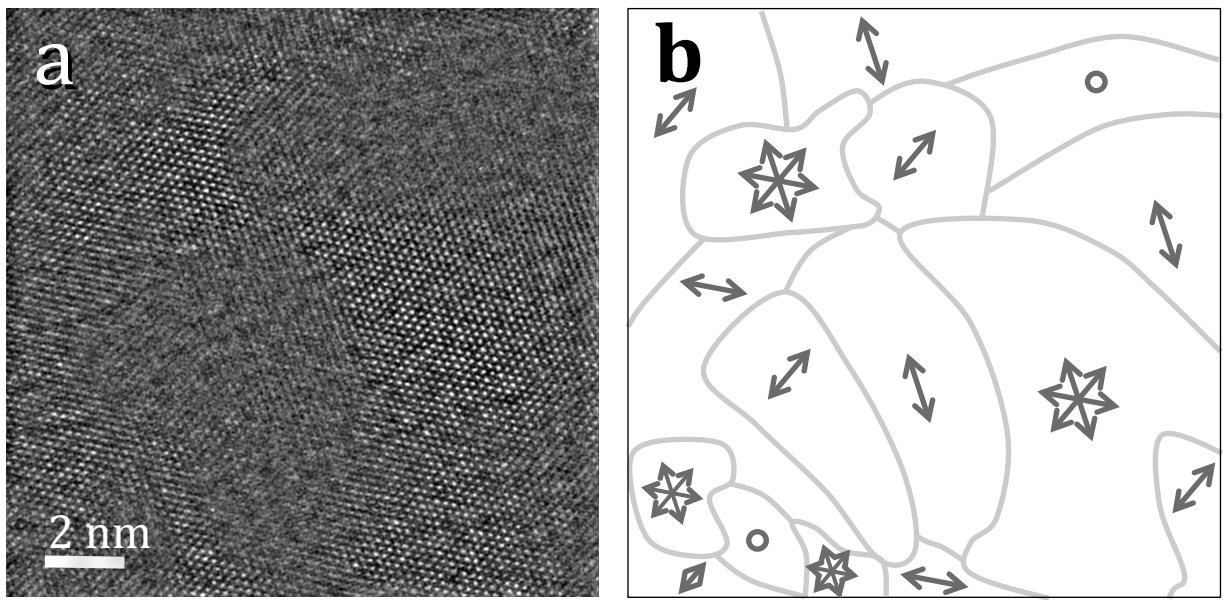

Fig 8: (a) High resolution HAADF-STEM micrograph of the Zr-Nb-Ti-Ta-Hf high-entropy alloy deformed at 200 °C up to 9 %, (b) schematic representation of the differently oriented areas in (a).

up to about 7 %, imaged using a (0 1 1) reflection. The dislocation density is significantly higher than in the undeformed material. Due to the very high dislocation density,

Burgers vector analysis or the quantification of the dislocation density is impossible by means of TEM. Fig. 7c is a micrograph of a sample deformed at 200 °C up to 9 %, imaged using a (0 1 1) reflection. The microstructure is considerably different than for the case of the sample deformed at RT. The overall image contrast is highly mottled and the dislocation density is lower than in the sample deformed at RT.

A high resolution STEM micrograph of the same sample is shown in Fig. 8a. The image, which was slightly FFT-filtered for high-frequency noise reduction, was taken in a sample area showing mottled contrast. It is seen that the microstructure is not uniform but diverged into areas of different orientation, which are schematically indicated in Fig. 8b. In the micrograph four areas are seen (marked by three crossed double-headed arrows in Fig. 8b) that are oriented along the [1 1 1] direction, which also corresponds to the globally aligned orientation of the sample. The micrograph however also contains diverse differently oriented areas. Some (marked by double headed arrows) are tilted along a single axis with respect to the global orientation, so that lattice fringes are seen. Areas showing three sets of fringes, which are contained in the [1 1 1] direction, are present. Two areas, labeled "O" have no specific orientation and therefore appear grey without any observable fringe contrast. The average extension of the areas is about 2 to 5 nm and the boundaries are not sharp. It should be noted that in the undeformed material, the structure shows homogeneous orientation throughout the single grains, i.e. over distances in the millimeter range.

## 4. Discussion

The mechanical properties of HEAs are the subject of numerous publications, particularly in the context of high-strength applications [1, 5, 16, 9, 11]. Indeed plastic deformation tests on HEAs frequently show high yield stresses and flow stresses. Zhou et al. [16] observe a yield stress of 1500 MPa at RT in compression experiments on equiatomic Al-Cr-Co-Fe-Ni HEAs. The yield stress can be further increased by addition of Ti. Similar results were obtained by Zhang et al. [9, 10]. Li et al. report a RT strength of 1300 MPa for equiatomic Al-Cu-Co-Cr-Fe-Ni [17]. Note that the samples that were investigated in these studies were reportedly polycrystalline and contained multiple phases.

The yield stresses observed in our experiments are significantly lower, amounting to about 850 MPa at RT and decreasing to about 600 MPa at 300 °C. This is a respectable RT strength, but far below commercial high-strength materials or bulk metallic glasses [18] and also significantly below the values reported for other HEAs. Therefore our experiments do not support the notion that high strength is a typical feature of HEAs. The samples used in our experiments were homogeneous and single phased, i.e. free of precipitates and foreign phases. Also, the sample dimensions used are comparable to the extension of single grains of our material. On average our samples contain 2 to 10 single grains and therefore we consider the effect of grain boundaries on the results as negligible. In this respect we consider the plastic behavior observed in our experiments as intrinsic to Zr-Nb-Ti-Ta-Hf HEAs.

The plasticity of HEAs more similar to ours was investigated by Senkov et al. [19]. These authors deformed equiatomic refractory metal HEAs Nb-Mo-Ta-W and V-Nb-Mo-Ta-W, being single phased body-centered cubic. The Nb-Mo-Ta-W alloy at RT shows a yield stress of 1058 MPa, which is also considerably higher than that found in the present study, and their sample failed at a plastic strain of about 2.0 %. The five component alloy V-Nb-Mo-Ta-W showed an even higher yield strength of 1246 MPa and failed at a plastic strain of about 1.5 %. For both alloys, deformation tests were carried out up to temperatures of 1600 °C, leading to a decrease of the yield stress (values of 405 and 477 MPa are found for the four- and five-component alloy) and an increase of the fracture strain.

Besides the included elements in the HEA of these authors, a major difference with respect to the present experiments lies in the microstructure. First, the grain size of their alloys is relatively small (200 and 80 μm for the four- and five-component alloy, respectively). This leads to a much higher number of grains per deformation sample, on average about $1.3 \cdot 10^4$ and $2.1 \cdot 10^5$ grains per sample for the four- and five-component alloy, respectively, in contrast to 2 to 10 grains per sample in our experiments. On top of that, the grains showed a dendritic substructure with a dendrite arm spacing of about 20 to 30 μm in both alloys. The considerable difference in microstructure, compared to the material considered in the present study, may account for the observed differences in mechanical behavior. Indeed, in the post-mortem analysis of their samples, Senkov et al. [19] relate the observed deformation behavior to the presence of the grain structure.

In the present study, the thermodynamic activation parameters of a HEA were measured for the first time. In particular the reported values of the activation volume and the activation enthalpy may be helpful for the identification of the deformation mechanism and for simulations in a later stage of research. By now, the activation volume may be compared with the average volume per atom $V_a$. With a lattice constant of 0.32 nm $V_a$ assumes about 0.08 nm$^3$, which is much smaller than the measured activation volumes ranging between 0.5 and 2 nm$^3$.

We only take the low temperature results into account, since the occurrence of structural changes during the deformation process impedes the evaluation of the data using Eqns. (1) and (2) [13]. Our experiments indicate that at higher temperatures indeed the structure is not constant: The stress-strain curve at 300 °C has a different overall shape than those at lower temperatures and temperature cycling to a higher temperature leads to an increase in flow stress. This strongly indicates that structural changes have taken place during the equilibration time, and straightforward evaluation of the data using Eqn. (2) would lead to a negative activation enthalpy which has to be considered unphysical.

The occurring structural changes are also reflected in the activation volume data. At 300 °C the determined values are much higher than the values obtained for 200 °C and below. Up to 200 °C structural changes apparently occur at a sufficiently low rate, so that the first 10 to 20 seconds of stress relaxation, which are used for the evaluation according to Eqn. (1), are still unaffected, so that the data points follow a consistent curve. At this stage of research we tend to attribute the higher values obtained at 300 °C rather to structural changes, taking place at such high rates that they affect the stress relaxation right from the beginning, than to a change in deformation mechanism. Similarly, by now we dismiss the data point obtained for the activation enthalpy at 200 °C. Temperature cycling involves an equilibration time of about 30 minutes, which allows structural changes to occur (see below). Surely, in this respect also the activation enthalpy data points at lower temperatures have to be regarded with some caution, but they are consistent within the experimental scatter and assume values that make sense with respect to the imposed strain rate.

Another indicator for structural changes are the overshoots that occur upon reloading after stress relaxations at temperatures of 60 °C and higher. All relaxations were carried out for two minutes, which shows that changes take place on the timescale of a few minutes. The evaluation of the overshoots (Fig. 6) indicates that structural changes take place faster with increasing temperature, and at 60 and 100 °C the changes evolve over the course of the experiment – the amount of overshoot increases continuously and seems to saturate towards high strains.

The TEM investigations indicate that the differences in stress-strain behavior at RT (at which no overshoots are observed) and higher temperatures (showing overshoots) are connected with changes in dislocation density. The TEM specimens were prepared from samples that were deformed up to 8 to 10 % of strain. At that stage, the samples were unloaded, taken out of the testing machine and quenched on a cold metal plate. This process takes about two minutes, and therefore the TEM micrographs reflect a state, which includes structural changes. Since the dislocation density in the samples deformed at temperatures above RT is much lower, we conclude that the structural changes involve the annihilation of dislocations and a subsequent buildup of mobile dislocation density upon reloading. The latter may involve a stress threshold [20], which is reflected in the overshooting flow stress. The high-resolution STEM micrographs (Fig. 8) clearly show in which respect the microstructure evolves in the course of plastic deformation: while the structure initially is uniform and homogeneous, it is diverged into areas of different orientation after deformation. These areas are congruently connected; no dislocations are present at their interfaces.

These conclusions on the structural changes are, however, rather preliminary. Deeper insight should be obtained by dedicated deformation experiments up to different strains with subsequent TEM characterization and by in-situ straining experiments in the transmission electron microscope at different temperatures.

Finally, let us provide some reasonable criticism on the common interpretation of HEA plasticity in terms of solid-solution strengthening. It is frequently argued that a solid-solution strengthening mechanism in an extreme form should be operative in HEAs. Yeh et al. [1], for example, write that due to the absence of a matrix element all atoms may be regarded as solute atoms, which should dramatically enhance the strength of HEAs.

Zhang et al [9] explicitly specify the amount of resulting solid-solution strengthening as proportional to $c^{1/2}$, where c is the concentration of solute atoms.

It should however be noticed that the situation is not that simple. First of all, the concentration of solute atoms exceeds the Friedel limit [22], which defines the limit of Friedel statistics, assuming that single obstacles are overcome one at a time. Under common assumptions the upper limit amounts to $c \approx 10^{-4}$, which is violated in HEAs not only if all atoms are considered as solute atoms influencing dislocation movement, but also if we make the weaker assumption that only one species should be taken into account. Friedel statistics would lead to a concentration dependence of the solid-solution strengthening effect with $c^{1/2}$ dependence. The Mott limit, defining the validity of Mott statistics, which assumes weaker obstacles at higher concentrations, ranges between 0.01 and 0.1, is also exceeded in ordinary HEAs. For Mott statistics, according to Nabarro et al. [21] a $c^{3/2}$ dependence (see also [14]) and, according to a later more detailed treatment by Arsenault, a $c^{1/2}$ dependence is expected [22].

The formal (and useful) definition of a HEA requires that none of the atomic species included constitutes a major component. The (more or less strict) requirement of equiatomic composition impedes the decision, what defines the matrix in which the solute atoms are solved. The closer a HEA fulfills the formal definition and formation criteria (requiring a narrow atomic size distribution [9]) the less obvious this becomes. Accordingly, it remains undefined with respect to which reference an increase in strength (be it proportional to $c^{1/2}$ or $c^{2/3}$) is effected.

The present authors feel that an adequate description of the plasticity of HEAs requires a new approach. Rather than defining a contrasting matrix and species of solute atoms, as required in a conventional solid-solution strengthening treatment [23], which is in conflict with the definition of HEAs, a scenario should be considered, in which the dislocations move through a strongly corrugated nonperiodic potential. The corrugations appear on a length scale comparable to or even smaller than the Burgers vector. To the authors' knowledge, no applicable theory is available as yet.

**Acknowledgment**

The authors thank Prof. Walter Steurer for communication of the single-phase composition of the presently investigated HEA, Marc Heggen for valuable discussions, Eva-Maria Würtz for sample preparation and characterization, and Maximilian Kruth for the preparation of TEM specimens.